\documentclass{eptcs}
\usepackage{comment}
\usepackage{amsmath}
\usepackage{amssymb}
\usepackage{amsthm}
\usepackage{url}
\usepackage{alltt}
\usepackage{stmaryrd}
\usepackage{graphicx}
\usepackage{subfigure}
\usepackage{tabularx}
\usepackage[all]{xy}

\newcommand{\cequal}{{\stackrel{ \rm e }{\; = \;}}}

\newcommand{\definedAs}{\:\,{\stackrel{{de\!f}}{=}}\:\,}

\newtheorem{lemma}{Lemma}
\newtheorem{theorem}{Theorem}
\newtheorem{corollary}{Corollary}

\newcommand{\mgu}{{\mathit{mgu}\;}}

\newcommand{\emptylist}{{[ \;]}}
\def\authorrunning{S. Kothari \& J. Caldwell}
\def\titlerunning{A Machine Checked Model of Idempotent MGU Axioms For Lists of Equational Constraints}
\begin{document}
\title{A Machine Checked Model of Idempotent MGU Axioms For Lists of Equational Constraints}
\author{Sunil Kothari,  James Caldwell\\
 Department of Computer Science,\\
  University of Wyoming, USA
}
\date{}
\maketitle
\begin{abstract}
We present formalized proofs verifying that the first-order
unification algorithm defined over lists of satisfiable constraints generates
a most general unifier (MGU), which also happens to be idempotent. All of our
proofs have been formalized in the Coq theorem prover.  Our proofs show that
finite maps produced by the unification algorithm provide a model of the
axioms characterizing idempotent MGUs of lists of constraints.  The axioms that
serve as the basis for our verification are derived from a standard set by
extending them to lists of constraints.  For us, constraints are equalities
between terms in the language of simple types.  Substitutions are formally
modeled as finite maps using the Coq library {\it
  Coq.FSets.FMapInterface}. Coq's method of functional induction is the main
proof technique used in proving many of the axioms.
\end{abstract}
\section{Introduction}
\label{sec:introduction}

We present formalized proofs verifying that the first-order
unification algorithm defined over lists of satisfiable constraints generates a
most general unifier (MGU), which also happens to be idempotent.  All of our
proofs have been formalized in the Coq theorem prover \cite{book:coq}.  Our
proofs show that substitutions produced by the unification algorithm provide a
model of the axioms characterizing the idempotent MGUs of lists of constraints.

The formalization and verification presented here was motivated by our work on
to verifying Wand's constraint based type inference algorithm \cite{paper:wand}
(and to verify our extension of Wand's algorithm to include polymorphic let
\cite{paper:kotcal1}).  
In the recent literature on machine certified proof of
correctness of type inference algorithms
\cite{paper:wcoq,paper:wisabelle,bookchapter:urbannipkow}, most general
unifiers are characterized by four axioms. 

Recall that $\tau$ and $\tau'$ (in some language) are {\em unifiable} if there
exists a substitution $\rho$ mapping variables to terms in the language such
that $\rho(\tau) = \rho(\tau')$. In such a case, $\rho$ is called a {\em
  unifier}. A unifier $\rho$ is a {\em most general unifier} if for any other
unifier $\rho''$ there is a substitution $\rho'$ such that $\rho \circ \rho' =
\rho''$.

We consider the MGU axioms given by
Nipkow and Urban \cite{bookchapter:urbannipkow}. Let $\rho, \rho', \rho''$
denote substitutions {\em {i.e.}} functions mapping type variables to terms,
constraints are of the form $\tau \cequal \tau'$ where $\tau$ and $\tau'$ are
simple types and the symbol ${\mathsf{FTV}}$ is overloaded to denote the free
type variables of substitutions, constraints and types and the
notation. Composition of substitutions\footnote{The reader should note that in
  this paper, composition of functions is characterized by the equation $(\rho \circ\rho')(x)
  = \rho'(\rho(x))$.}  is denoted $\rho\circ\rho'$.  With these notational
conventions in mind, the MGU axioms are presented as follows:
\begin{small}
\[\begin{array}{cl}
(i) &  \mgu \rho \: (\tau_1 \cequal \tau_2)  \Rightarrow \rho(\tau_1)=\rho(\tau_2)\\
(ii) &  \mgu \rho \: (\tau_1 \cequal \tau_2) \; \wedge \; \rho'(\tau_1)=\rho'(\tau_2) \Rightarrow \exists \rho''. \rho' = \rho \circ \rho''\\
(iii) &  \mgu \rho \: (\tau_1 \cequal \tau_2)  \Rightarrow \mathsf{FTV}(\rho) \subseteq \mathsf{FTV}(\tau_1 \cequal \tau_2)\\
(iv) &   \rho(\tau_1) = \rho(\tau_2) \Rightarrow \exists \rho'. \;\mgu \rho'\: (\tau_1 \cequal \tau_2)
\end{array}\]
\end{small}

These axioms, modeling MGUs, have proved useful in verifying
substitution-based type inference algorithms where the constraints are solved
as they are generated, one at a time. In constraint-based type inference
algorithms like Wand's, the constraints are generated before they are
solved. Thus, for use in the constraint based setting, we lift the MGU axioms
to lists of constraints. To do so, we restate the standard axioms to apply to
constraint lists, add two new axioms which characterize MGUs of lists of
constraints; one axiom for the empty list and another for lists constructed by
appends.  Also, reasoning about Wand's type inference algorithm requires the
MGUs be idempotent, so we add another axiom for idempotency. Idempotent MGUs
have the nice property that their domain and range elements are disjoint.

We proceed by characterizing idempotent MGUs for lists of equational
constraints by presenting seven axioms. Then we show that the first order
unification algorithm models those axioms. The theorems and supporting lemmas
mentioned in this paper have been formalized and verified in Coq
\cite{manual:coq} - a theorem prover based on calculus of inductive
constructions \cite{paper:coc}. In the formalization, we represent
substitutions using Coq's finite map library \cite{manual:coq_fmapinterface}.

To start, we generalize the standard MGU axioms to constraint lists. In
addition to the notations introduced above, if $C$ is a list of constraints,
$\rho\models{}C$ (read $\rho$ {\em{satisfies}} $C$) means that $\rho$ unifies
all constrains in $C$. Let $C$ denote a constraint list, then the MGU axioms
(for a list of constraints) are:\\

\begin{tabular}{cl}
$\mbox{\hspace{1cm}}(i)$ & $\mgu \rho \; C \; \Rightarrow \rho \models C$\\
$\mbox{\hspace{1cm}}(ii)$ &  $\mgu \rho \;C \; \wedge\; \rho' \models C \Rightarrow \exists \rho''. \;\rho' = \rho \circ \rho''$\\
$\mbox{\hspace{1cm}}(iii)$ &  $\mgu \rho \;C \; \Rightarrow \mathsf{FTV}\;(\rho) \subseteq \mathsf{FTV}\;(C)$\\
$\mbox{\hspace{1cm}}(iv)$ &   $\rho \models \;C \;\Rightarrow \exists \rho'. \;\mgu \rho'\: C$
\end{tabular}\\\\


\noindent{}To the axioms just mentioned we add three more axioms that characterize
idempotent MGUs for a list of equational constraints.  List append is denoted
by $\mathit{++}$.

\begin{tabular}{cl}
$\mbox{\hspace{1cm}}(v)$ &  $\mgu \rho \; C \; \Rightarrow \;\rho \circ \rho =  \rho$\\
$\mbox{\hspace{1cm}}(vi)$  & $\mgu \rho \;[\;] \Rightarrow \rho = Id$\\
$\mbox{\hspace{1cm}}(vii)$ &  $\mgu \rho' \;C'\;  \wedge \;\mgu \rho'' \: (\rho'(C'')) \wedge \mgu \rho \;(C' \; \mathit{++}\; C'')\Rightarrow \rho = \rho' \circ \rho''$\\\\
\end{tabular}

\noindent
These additional axioms are mentioned elsewhere in the unification literature,
namely \cite{paper:elmar, paper:LMM}. The statement of axiom {\it{vii}} is
convenient in proofs where constraint lists are constructed by combining lists
of constraints rather than adding them one at a time.  A lemma characterizing
lists constructed by conses is easily proved from this axiom.


Formalizing substitutions as finite maps in Coq, we show that first-order
unification (${\mathsf{unify}}$) is a model of the MGU axioms. To distinguish
the formal representation of substitutions as finite maps from mathematical
functions, we denote finite maps by $\sigma$, $\sigma'$, $\sigma_1$,
etc. Mathematical functions enjoy extensional equality while finite maps do not
(more about this later).  We write $ \rho \approx \rho'$ to denote extensional
equality for finite maps; {\em{i.e.}} that under application they agree
pointwise on all inputs.  With these considerations in mind, we have proved the
following in Coq:\\

\begin{tabular}{cl}
$\mbox{\hspace{1cm}}(i)$ &   $\mathsf{unify}(C) = \sigma \;  \Rightarrow \sigma \models C$\\
$\mbox{\hspace{1cm}}(ii)$ & $(\mathsf{unify}(C) = \sigma \;  \wedge \sigma' \models C) \Rightarrow \exists \sigma''. \;\sigma' \approx \sigma \circ \sigma''$\\ 
$\mbox{\hspace{1cm}}(iii)$ &  $\mathsf{unify}(C) = \sigma \;  \Rightarrow \mathsf{FTV}(\sigma) \subseteq \mathsf{FTV}(C)$\\
$\mbox{\hspace{1cm}}(iv)$ &  $\sigma \models  C  \;  \Rightarrow \exists \sigma'. \;\mathsf{unify}(C) =  \sigma'$\\
$\mbox{\hspace{1cm}}(v)$ &   $\mathsf{unify}(C) =  \sigma \;  \Rightarrow \sigma \circ \sigma  \approx \sigma$\\
$\mbox{\hspace{1cm}}(vi)$ &   $\mathsf{unify}([\;]) =   \sigma \;  \Rightarrow \sigma = \sigma_\mathbb{E}$\\
$\mbox{\hspace{1cm}}(vii)$ & $(\mathsf{unify}(C') = \sigma' \;  \wedge \mathsf{unify}(\sigma'(C'')) =  \sigma'' \wedge \mathsf{unify}(C' \; {++}\; C'') =  \sigma) \Rightarrow  \;\sigma \approx \sigma' \circ \sigma''$\\\\
\end{tabular}


The rest of this paper is organized as follows: Section \ref{sec:typesandsubst}
introduces a number of formal definitions and terminologies needed for this
paper. It also includes more discussion about substitutions represented as
finite functions.  Section \ref{sec:unification} describes the formalization of
a first-order unification algorithm and the termination argument. Section
\ref{sec:mgus} describes the functional induction tactic and the theorems and
lemmas proved in the verification that {\sf{unify}} models the idempotent MGU
axioms.  Finally, Section \ref{sec:conclusions} mentions related work and also
summarizes our current work.
\section{Types and Substitutions}
\label{sec:typesandsubst}
Unification is implemented here over a language of simple types
given by the following grammar:\\
\begin{small}
   \begin {tabular}{lcl}
  \mbox{\hspace{2cm}}  $\tau$   ::= & $\alpha$  $\mid$ & $ \tau_1 \rightarrow \tau_2 $ \\
\end{tabular}\\
  {\mbox{\hspace{8em}}} where $\alpha$ is a type variable, and $\tau_1,\tau_2\in\tau$ are type terms. \\
\end{small}
\noindent
Thus, a type is either a type variable or a function type.
We define the list of \emph{free\footnote{Strictly speaking, since we have no
    binding operators in the language of simple types the modifier ``free'' is
    unnecessary, we include it here anticipating a more complex language of
    types in future developments.} variables of a type} ($\mathsf{FTV}$) as:\\
\begin{small}
   \begin{tabular}{lcl}
     $\mbox{\hspace{2cm}}\mathsf{FTV}(\alpha)$ & $\definedAs$ & $[\alpha]$\\
     $\mbox{\hspace{2cm}}\mathsf{FTV}(\tau \rightarrow \tau')$ & $\definedAs$ & $\mathsf{FTV}(\tau)\,\, \mathrm{++}\,\, \mathsf{FTV}(\tau')$ 
\end{tabular}
\end{small}

We also have equational constraints of the form  $\tau \cequal \tau'$, where $\tau, \tau'$ are types.
The list of \emph{free variables of a constraint list}, also denoted by $\mathsf{FTV}$, is given as: \\
\begin{small}
\begin{tabular}{lcl}
     $\mbox{\hspace{2cm}}\mathsf{FTV}(\emptylist)$ & $\definedAs$ &  $\emptylist$\\
     $\mbox{\hspace{2cm}}\mathsf{FTV}((\tau_1 \cequal \tau_2)::C)$ &$\definedAs$ & $\mathsf{FTV}(\tau_1)\,\, \mathit{++}\,\,\mathsf{FTV}(\tau_2) \,\,\mathit{++}\,\, \mathsf{FTV}(C)$
\end{tabular}
\end{small} \vspace{.125em}

Substitutions are formally represented as finite maps where the domain of the
map is the collection of type variables and the codomain is the simple
types. Application of a finite map to a type is defined as:\\
$\begin{array}{lcl} \mbox{\hspace{2cm}}\sigma(\alpha) &\definedAs & \left\{
\begin{array}{cl} \tau& \mathit{if}\; \langle{}\alpha,\tau \rangle \;\in\;
\sigma\; \\ \alpha & \mathit{otherwise}
                                         \end{array} \right.\\
\mbox{\hspace{2cm}}\sigma(\tau_1 \rightarrow \tau_2) & \definedAs &  \sigma (\tau_1) \rightarrow \sigma (\tau_2)
\end{array}$

\noindent
Application of a finite map to a constraint is defined similarly as:\\
\begin{small}
\begin{tabular}{lcl}
  $\mbox{\hspace{2cm}}\sigma (\tau_1 \cequal \tau_2)$ &$\definedAs$& $\sigma(\tau_1) \cequal \sigma(\tau_2)$\\
\end{tabular}
\end{small}

\noindent
Since Coq's finite maps are not extensional, we define extensionality 
($\approx$) as a relation on finite maps as follows:\\
\begin{small}
\begin{tabular}{lcl}
$\mbox{\hspace{2cm}}\sigma \approx \sigma'$ &$\definedAs$& $\forall{}\alpha.\: \sigma(\alpha) = \sigma'(\alpha)$
\end{tabular}
\end{small}

\noindent
Moreover, the equality can be extended to all types as given by the following lemma:
\begin{lemma}\label{lemma:squiggle_ext_lift}
$\forall{}\alpha.\: \sigma(\alpha) = \sigma'(\alpha)  \Leftrightarrow \forall{}\tau.\: \sigma(\tau) = \sigma'(\tau) $ \vspace{-.45em}
\end{lemma}


\subsection{Implementing Substitutions as Finite Maps}
The representation of substitutions and the libraries available to a user plays
a very important role in the formalization. In the verification literature,
substitutions have been represented as functions
\cite{bookchapter:urbannipkow}, as lists of pairs \cite{paper:wcoq}, and as
sets of pairs \cite{paper:paulsonLCF}. We represent substitutions as finite
functions (a.k.a finite maps in Coq).  We use the Coq finite map library
{\it{Coq.FSets.FMapInterface}} \cite{manual:coq_fmapinterface}, which provides
an axiomatic presentation of finite maps and a number of supporting
implementations.  However, it does not provide an induction principle for
finite maps, and forward reasoning is often needed to use the library. We found
we did not need induction to reason on finite maps, though there are natural
induction principles we might have proved
\cite{paper:collins_syme_95,
book:manna_waldinger_85}. The fact that the library
does not provide for extensional equality of finite maps means that, for
example, the following simple lemma does not hold:
\begin{lemma}
 $\sigma_\mathbb{E} \circ \sigma_\mathbb{E} = \sigma_\mathbb{E}$
\end{lemma}

\noindent
But the following is easily proved:
\begin{lemma}
 $\forall \tau. (\sigma_\mathbb{E} \circ \sigma_\mathbb{E})(\tau)= \sigma_\mathbb{E}(\tau)$
\end{lemma}

To give a feel of the Coq's finite map library, we define free type variables
of a substitution, and the substitution composition operator using the finite
map library functions. In the definitions below, we follow Coq's namespace
conventions; every library function has a qualifier which denotes the library
it belongs to. For example, $\mathit{M.map}$ is a function from the finite maps
library (${\mathit{M}}$) which maps a function over the range elements of a
finite map, whereas $\mathit{List.map}$ is a function from the list library.

\noindent
First, we define the list of free type variables of a substitution:\\
\begin{tabular}{lcl}
$\mbox{\hspace{2cm}}\mathsf{FTV}(\sigma)$ &$\definedAs$ & $\mathsf{dom\_subst}(\sigma) \,\,\mathit{++} \,\,\mathsf{range\_subst}( \sigma)$\\
\end{tabular}

To consider the {\em domain} and {\em range} elements of a finite function (and this is the key
feature of the function being finite), we use the finite map library function
$\mathsf{M.elements}$.  $\mathsf{M.elements}(\sigma)$ returns a list of pairs (key-value pairs)
corresponding to the finite map $\sigma$. The domain and range elements of a
substitution are defined as:\\
\begin{tabular}{lcl}
$\mbox{\hspace{2cm}}\mathsf{dom\_subst}(\sigma)$ & $\definedAs$& ${\mathsf{List.map}\;  (\lambda t. \mathsf{fst}\;(t)) \;(\mathsf{M.elements} (\: \sigma))}$\\
$\mbox{\hspace{2cm}}\mathsf{range\_subst}(\sigma)$ &  $\definedAs$& ${\mathsf{List.flat\_map}\;  (\lambda t.\mathsf{FTV}\; (\mathsf{snd}\; (t)))\; (\mathsf{M.elements} (\: \sigma))}$\\
\end{tabular}

\noindent
The function $\mathsf{List.flat\_map}$ is also known as $\mathsf{mapcan}$ in LISP and $\mathsf{concatMap}$ in Haskell.
Next, we define a few utility functions to help us define the  composition operator $\circ$.  Applying a substitution $\sigma'$ to a substitution $\sigma$ means applying $\sigma'$ to the range elements of $\sigma$.\\
\begin{tabular}{lcl}
$\mbox{\hspace{2cm}}\sigma'(\sigma)$ & $\definedAs$ & ${\mathsf{M.map}}\; (\lambda \tau. \sigma'(\tau)) \; \sigma$
\end{tabular}

\noindent
The function $\mathsf{subst\_diff}$ is used to define composition of finite maps, and is defined as:\\
\begin{tabular}{lcl}
$\mbox{\hspace{2cm}}\mathsf{subst\_diff}\: \sigma\: \sigma'$ & $\definedAs$& ${\mathsf{M.map2}}\; \mathsf{choose\_subst}\; \sigma \; \sigma'$\\
\end{tabular}

\noindent
In this definition, $\mathsf{M.map2}$ is defined in Coq library as the function
that takes two maps $\sigma$ and $\sigma'$, and creates a map whose binding
belongs to either $\sigma$ or $\sigma'$ based on the function
$\mathsf{choose\_subst}$, which determines the presence and value for a key
(absence of a value is denoted by $\mathsf{None}$). The values in the first map
are preferred over the values in the second map for a particular key.  The
function $\mathsf{choose\_subst}$ is defined as:\\
\begin{tabular}{lcl}
$\mbox{\hspace{2cm}}\mathsf{choose\_subst}\; (\mathsf{Some}\,\tau_1) \;\; (\mathsf{Some} \,\tau_2)$& $\definedAs$ &$\mathsf{Some}\; \tau_1$\\
$\mbox{\hspace{2cm}}\mathsf{choose\_subst}\; (\mathsf{Some}\,\tau_1)\;\; \mathsf{None}$ &$\definedAs$& $\mathsf{Some}\; \tau_1$\\
$\mbox{\hspace{2cm}}\mathsf{choose\_subst}\; \mathsf{None} \;\; (\mathsf{Some}\, \tau_2)$&$\definedAs$& $\mathsf{Some}\, \tau_2$\\
$\mbox{\hspace{2cm}}\mathsf{choose\_subst}\; \mathsf{None}\;\; \mathsf{None}$ &$\definedAs$& $\mathsf{None}$
\end{tabular}

\noindent
Finally, the composition of finite maps ($\circ$) is defined as: \\
\begin{tabular}{lcl}
$\mbox{\hspace{2cm}}\sigma \circ \sigma'$ & $\definedAs$ &  $\mathsf{subst\_diff}\,\, \sigma'(\sigma) \,\,\,\sigma'$ 
\end{tabular}

\noindent
Substitution composition application to a type has the following property:
\begin{theorem}\label{thm:composition_apply}
$\forall \sigma.\; \forall \sigma'. \; \forall{}\tau.(\sigma\circ\sigma')(\tau) = \sigma'(\sigma(\tau))$
\end{theorem}
\begin{proof}
By induction on the type $\tau$ followed by case analysis on the binding's occurrence in the composed substitution and in the individual substitutions.
\end{proof}
\noindent{}Interestingly, the base case (when $\tau$ is a type variable) is
more difficult than the inductive case (when $\tau$ is a compound
type). Incidentally, the same theorem has been formalized in Coq
\cite{paper:wcoq}, where substitutions are represented as lists of pairs, but
the proof there required 600 proof steps. We proved Theorem
\ref{thm:composition_apply} in about 100 proof steps.

\section{First-Order Unification}
\label{sec:unification}
 We use the following standard presentation of the first-order unification algorithm:\\
\begin{small}
\begin{tabular}{lll}
$\mathsf{unify} \: \emptylist$  & $\definedAs$ &  $Id$ \\
$\mathsf{unify}\:((\alpha \cequal \beta):: C)$& $\definedAs$ &  \sf{if} $\alpha = \beta$ \sf{then} $\mathsf{unify}(C)$ \sf{else}  $\{\alpha \mapsto \;\beta \} \circ \mathsf{unify}\:(\{\alpha \mapsto \beta\}(C))$ \\
$\mathsf{unify}\:((\alpha \cequal \tau) :: C)$ & $\definedAs$   & \sf{if} $\alpha$ occurs in $\tau$ \sf{then} Fail \sf{else} 
  $\{\alpha \mapsto \;\tau \} \circ \mathsf{unify}\; (\{\alpha\mapsto \tau\}(C))$\\
 $\mathsf{unify}((\tau \cequal \alpha) :: C)$& $\definedAs$ &  \sf{if} $\alpha$ occurs in $\tau$ \sf{then} Fail \sf{else} 
   $\{\alpha \mapsto \;\tau \} \circ \mathsf{unify}\: (\{\alpha\mapsto \tau\}(C))$\\
 $\mathsf{unify}\:((\tau_1 \rightarrow \tau_2 \cequal \tau_3 \rightarrow \tau_4):: C)$ & $\definedAs$  &  $\mathsf{unify}((\tau_1 \cequal \tau_3):: (\tau_2 \cequal \tau_4) :: C)$  \\\\
\end{tabular}
\end{small}

\noindent
This specification is written in a functional style.  It would also have been
possible to formalize {\sf{unify}} in a relational style. A discussion of the
trade-offs between these two styles of formalization Coq can be found in
\cite{paper:func_ind_orig}.  Since Coq's type theory requires functions to be
total, the functional style carries an overhead; we need a value to represent
failure. We used Coq's {\tt option} type to make first-order unification total.
The {\em option} type ({\em maybe} in Haskell) is defined in Coq as follows:\\
$\mbox{\hspace{2cm}}\mathsf{Inductive\; option\; (A:Set)\; : \; Set \; := \;
Some \; (\_:A) \; |\; None.}$

\noindent
The constructor $\mathsf{None}$ indicates failure and the term
$\mathsf{Some}(\sigma)$ indicates success (with $\sigma$ as the result).  In
the presentation here, we omit the $\mathsf{None}$ and $\mathsf{Some}$
constructors.  In virtually all theorems proved here, the $\mathsf{None}$ case
is trivial.

The presentation of the unification algorithm given here is general recursive,
{\em i.e.}, the recursive call is not necessarily on a structurally smaller
argument. Various papers have discussed the non-structural recursion used in
the standard first-order unification algorithm. McBride has given a
structurally recursive unification algorithm \cite{ paper:mcbride}. Bove
\cite{paper:bove} gives an algorithm similar to ours and proves termination in
Alf \cite{manual:alf}. We believe our presentation of the algorithm is more
perspicuous than Bove's although a similar termination argument works here. To
allow Coq to accept our definition of unification, we have to either give a
measure that shows that recursive argument is smaller or give a well-founded
ordering relation. We chose the latter.  We use the standard lexicographic
ordering on the triple: $< \mid\! C_{FVC} \!\mid, \mid\!  C_\rightarrow \!\mid,
\mid\! C \!\mid>$, where\\
 \begin{small}
 $\mbox{\hspace{1cm}}\mid\! C_{FVC}\! \mid$ is the number of {\em unique} free variables in a constraint list;\\
 $\mbox{\hspace{1cm}}\mid \!C_\rightarrow \!\mid$ is the total number of arrows  in the constraint list;\\
 $\mbox{\hspace{1cm}}\mid\!C \!\mid$ is the length of the constraint list.\\
\end{small}
Our triple is similar to the triple proposed by others \cite{paper:bove, paper:baadersnyder, book:apt}, but a little simpler.

\begin{table}
\begin{small}
\begin{tabular}{|l|l|l|c|c|c|}
\hline
Original call & Recursive call & Conditions, if any  & $\mid\! C_{FVC}\! \mid$ & $\mid \!C_\rightarrow\! \mid$ & $\mid\! C \!\mid$ \\
\hline
 $(\alpha \cequal \alpha)::C$ &  $C$ &$\alpha \in \mathsf{FVC}(C)$ & - & - & $\downarrow$ \\  
 $(\alpha \cequal \alpha)::C$ &  $C$&$\alpha \notin \mathsf{FVC}(C)$ & $\downarrow$ & -  & $\downarrow$  \\
 $(\alpha \cequal \beta)::C$ & \; $\{\alpha\mapsto \beta\}(C)$&$ \alpha \neq \beta $ & $\downarrow$ & - & $\downarrow$ \\  
 $(\alpha \cequal \tau)::C$ & \; $\{\alpha\mapsto \tau\}(C)$&$ \alpha \notin \mathsf{FTV}(\tau)$ $\wedge$ $ \alpha \notin \mathsf{FVC}(C)$ & $\downarrow$ & $\downarrow$  & $\downarrow$ \\
 $(\alpha \cequal \tau)::C$ & \; $\{\alpha\mapsto \tau\}(C)$&$ \alpha \notin (\mathsf{FTV}(\tau)$ $\wedge$ $ \alpha \in \mathsf{FVC}(C)$ & $\downarrow$ &  $\uparrow$ & $\downarrow$ \\
 $(\tau \cequal \alpha)::C$ & \; $\{\alpha\mapsto\tau\}(C)$&$ \alpha \notin \mathsf{FTV}(\tau)$ $\wedge$ $ \alpha \notin \mathsf{FVC}(C)$ & $\downarrow$ & $\downarrow$  & $\downarrow$ \\
 $(\tau \cequal \alpha)::C$ & \; $\{\alpha\mapsto\tau\}(C)$&$ \alpha \notin \mathsf{FTV}(\tau )$ $\wedge$ $ \alpha \in \mathsf{FVC}(C)$ & $\downarrow$ &  $\uparrow$ & $\downarrow$ \\
 $((\tau_1 \rightarrow \tau_2)$  &   $((\tau_1 \cequal \tau_3)$& None & - &$\downarrow$  &$\uparrow$ \\
$\mbox{\hspace{0.25cm}}\cequal (\tau_3 \rightarrow \tau_4))::C$& {\mbox{\hspace{1em}}}$::( \tau_2 \cequal \tau_4) :: C)$ &&&&\\
\hline
\end{tabular}
\end{small}
\caption{Properties of the termination measure components on the recursive call}
\label{table:unifitermination}
\end{table}
Table \ref{table:unifitermination} shows how these components vary depending on
the constraint at the head of the constraint list. The table closely follows
the reasoning we used to satisfy the proof obligations generated by the above
specification \cite{paper:unif09}.  We use -, $\uparrow$, $\downarrow$ to
denote whether the component is unchanged, increased or decreased,
respectively.  We might have used finite sets here (for counting the unique
free variables of a constraint list), but we used lists because of our
familiarity with the list library. We found the existing Coq list library offers
excellent support for reasoning about lists in general, and unique lists in
particular. Coq also provides a library to reason about sets as lists modulo
permutation.  

We found the following lemma mentioned in the formalization of Sudoku puzzles
by Laurent Th\'ery \cite{paper:sudoku} very useful in our termination proofs.
\begin{lemma}\label{lemma:ulist_incl_length_strict}
\[\forall{} l, l' :list\; D,\; \mathsf{NoDup}\; l\; \Rightarrow \mathsf{NoDup}\; l' \Rightarrow \mathsf{List.incl}\; l\; l' \Rightarrow \neg \mathsf{List.incl}\; l'\; l \Rightarrow (\mathsf{List.length}\; l)\;< (\mathsf{List.length}\; l')\]
\end{lemma}
\noindent{}This lemma nicely relates list inclusion to length.

\section{Verification of the Model}
\label{sec:mgus}
Now we present the proofs of the theorems verifying our model of the idempotent
MGU axioms. The underlying theme in almost all of the proofs presented below is
the use of the $\mathsf{functional\; induction}$ tactic
\cite{paper:func_ind_orig} in Coq. This tactic is available to us because we
have specified first-order unification in a functional style rather than the
relational style. The functional induction technique generates an induction
principle for definitions defined using the $\mathsf{Function}$ keyword.  Given
a general recursive algorithm known to terminate (termination requires a
separate proof), the induction principle generated for that particular algorithm
allows a symbolic unfolding of the computation with induction hypotheses for all
recursive calls.
This technique is featured in other theorem provers and was pioneered in
Nqthm by Boyer and Moore \cite{book:boyermoore}.

Functional induction is obviously stronger than the normal list induction, it
closely follows the syntax of the definition and tends to generate induction
hypotheses of exactly the right form needed.  The actual induction principle is
available in \cite{paper:unif09}. The induction principle for the unification
algorithm itself is rather long because of the number of cases involved; there
are five cases - three of which have three sub-cases each. 

In the next few sections, we present the formal statements of the most
important lemmas involved in the proofs of each of the axioms. For many of
these lemmas, we describe the main technique involved in the proofs. Due to
limitations on space, lemmas stated without comment on their proofs should be
assumed to follow by structural induction on a constraint list or type.

\subsection{Axiom i}

\begin{lemma}\label{lemma:satisfy_and_compose_subst}
$\forall{} \alpha. \: \forall{}C.\;\forall{} \sigma.\; \forall \tau. \; \sigma  \models \{\alpha \mapsto \tau\}(C) \Rightarrow 
  (\{\alpha \mapsto \tau\}\circ \sigma) \models C $
\end{lemma}



\begin{theorem}
 $\forall{} C. \;\forall{}\sigma.\; \mathsf{unify}(C) = \sigma \:  \Rightarrow \sigma \models C$
\end{theorem}

\begin{proof}
Choose an arbitrary $C$. By functional induction on $\mathsf{unify}\; C$, there are two main cases:
\begin{description}
\item Case $C = [\;]$. Follows trivially since any substitution satisfies an empty constraint list.
\item Case $C \neq [\;]$. 
We consider the various cases based on the constraint at the head of the constraint list.
\begin{enumerate}
\item Case $(\alpha \cequal \alpha)::C'$. This case follows from the induction hypothesis.
\item Case $(\alpha \cequal \beta)::C'$ and $ \alpha \neq \beta$. The reasoning is similar to case 3 below.
\item Case $(\alpha \cequal \tau_1 \rightarrow \tau_2)::C'$ and $\alpha \notin \mathsf{FTV}(\tau_1 \rightarrow \tau_2)$.
We know $\mathsf{unify} (\{\alpha \mapsto \tau_1 \rightarrow \tau_2\}(C')) = \sigma'$ and 
the induction hypothesis is\\
 $\forall \sigma.\: \mathsf{unify} (\{\alpha \mapsto \tau_1 \rightarrow \tau_2\}(C')) = \sigma $
 $\Rightarrow \sigma \models \{\alpha \mapsto \tau_1 \rightarrow \tau_2\}(C')$.\\ 
 We have to show \\
$\forall \sigma. \sigma = (\{\alpha \mapsto \tau_1 \rightarrow \tau_2\} \circ \sigma')\Rightarrow \sigma  \models (\alpha \cequal \tau_1 \rightarrow \tau_2)::C'$.
Pick an arbitrary $\sigma$. Assume $\sigma = \{\alpha \mapsto \tau_1 \rightarrow \tau_2\} \circ \sigma'$. We must show 
$(\{\alpha \mapsto \tau_1 \rightarrow \tau_2\} \circ \sigma') \models (\alpha \cequal \tau_1 \rightarrow \tau_2)::C'$.
Since we know $\mathsf{unify} (\{\alpha \mapsto \tau_1 \rightarrow \tau_2\}(C')) = \sigma'$, so by the induction hypothesis we know
 $\sigma' \models \{\alpha \mapsto \tau_1 \rightarrow \tau_2\}(C')$. We must show 
$(\{\alpha \mapsto \tau_1 \rightarrow \tau_2\} \circ \sigma') \models (\alpha \cequal \tau_1 \rightarrow \tau_2)::C'$. By the definition of satisfiability, we must show:
\begin{enumerate}
\item  $(\{\alpha \mapsto \tau_1 \rightarrow \tau_2\} \circ \sigma') \models (\alpha \cequal \tau_1 \rightarrow \tau_2)$.\\
 By Theorem  \ref{thm:composition_apply} and the definition of satisfiability, we must show $\sigma' (\{\alpha \mapsto \tau_1 \rightarrow \tau_2\}(\alpha))$
= $\sigma' (\{\alpha \mapsto \tau_1 \rightarrow \tau_2\}(\tau_1 \rightarrow \tau_2))$. 
  Since we know $\alpha \notin \mathsf{FTV} (\tau_1 \rightarrow \tau_2)$, so
  $\{\alpha \mapsto \tau_1 \rightarrow \tau_2\}(\tau_1 \rightarrow \tau_2) = \tau_1 \rightarrow \tau_2$ and the proof follows.
  
\item $(\{\alpha \mapsto \tau_1 \rightarrow \tau_2\} \circ \sigma') \models C'$. \\
Since we know $\sigma' \models \{\alpha \mapsto \tau_1 \rightarrow \tau_2\}(C')$, so by Lemma \ref{lemma:satisfy_and_compose_subst} we know  $(\{\alpha \mapsto \tau_1 \rightarrow \tau_2\} \circ \sigma') \models C'$ as was to be shown.
\end{enumerate}

\item Case $(\tau_1 \rightarrow \tau_2 \cequal \alpha)::C'$ and $\alpha \notin \mathsf{FTV}(\tau_1 \rightarrow \tau_2)$.
Same as case 3 above.
\item Case $(\tau_1 \rightarrow  \tau_2 \cequal \tau_3 \rightarrow \tau_4 )::C'$.  The induction hypothesis is \\
$\forall \sigma.\;  \mathsf{unify}((\tau_1 \cequal  \tau_3) :: (\tau_2 \cequal \tau_4)::C') = \sigma \Rightarrow 
 \sigma \models ((\tau_1 \cequal  \tau_3) :: (\tau_2 \cequal \tau_4)::C')$. 

We have to show \\
$\forall \sigma'.\;  \mathsf{unify}((\tau_1 \cequal  \tau_3) :: (\tau_2 \cequal \tau_4)::C') = \sigma' \Rightarrow 
 \sigma' \models ((\tau_1 \rightarrow  \tau_2 \cequal \tau_3 \rightarrow \tau_4 )::C')$. \\
Pick an arbitrary $\sigma'$ and assume $\mathsf{unify} \;((\tau_1 \cequal  \tau_3) :: (\tau_2 \cequal \tau_4)::C') = \sigma'$. Since we know $\mathsf{unify} \;((\tau_1 \cequal  \tau_3) :: (\tau_2 \cequal \tau_4)::C') = \sigma'$, so by the induction hypothesis we know \\ 
$\sigma' \models ((\tau_1 \cequal  \tau_3) :: (\tau_2 \cequal \tau_4)::C')$. But by the definition of satisfiability, we know \\
$\sigma'(\tau_1) =\sigma'(\tau_3)$, $\sigma'(\tau_2) = \sigma'(\tau_4)$ and $\sigma' \models C'$.  \\
To show  $\sigma' \models ((\tau_1 \rightarrow  \tau_2 \cequal \tau_3 \rightarrow \tau_4 )::C')$, we must show:
\begin{enumerate}
\item $\sigma' \models \tau_1 \rightarrow  \tau_2 \cequal \tau_3 \rightarrow \tau_4$. By the definition of satisfiability, we must show \\
$\sigma' (\tau_1 \rightarrow  \tau_2) = \sigma' (\tau_3 \rightarrow \tau_4)$. But we assumed $\sigma'(\tau_1) =\sigma'(\tau_3)$ and  $\sigma'(\tau_2) = \sigma'(\tau_4)$, so this case holds.
\item $\sigma' \models C'$. But that we already know.
\end{enumerate} 
\end{enumerate}
\end{description}
\end{proof}

\subsection{Axiom ii}

\begin{lemma}\label{lemma:constraint_satisfaction_and_substitution_instance}
$\forall C. \:\forall{}\sigma.\: \forall{}\alpha.\:\forall \tau.\: ( \sigma \models C\; \wedge \;\alpha \notin \mathsf{FTV}(\tau) \; \wedge\; \sigma(\alpha) = \sigma(\tau))  \:\Rightarrow \: \sigma \models \{\alpha \mapsto \tau\}(C)$
\end{lemma}
\begin{proof}
By induction on the constraint list $C$, followed by induction on the structure of the type $\tau$.
\end{proof}

\begin{theorem}
$\forall{} C.\;\forall{}\sigma. \forall \sigma'.\; (\mathsf{unify}(C) = \sigma \: \wedge \: \sigma' \models C) \Rightarrow \exists \sigma''. \;\sigma' \approx \sigma \circ \sigma''$ 
\end{theorem}
\begin{proof}
Choose an arbitrary constraint list $C$. By the definition of extensional
equality on finite maps, we must show $\forall{}\sigma. \forall \sigma'.\;
(\mathsf{unify}(C) = \sigma \: \wedge \: \sigma' \models C) \Rightarrow \exists
\sigma''. \;\forall \alpha.\; \sigma' (\alpha) = (\sigma \circ
\sigma'')(\alpha)$. \\By functional induction on $\mathsf{unify}(C)$, there are
two main cases:
\begin{description}
\item Case $C = [\;]$. Choose an arbitrary $\sigma$ and $\sigma'$. Assume $\mathsf{unify}([\;]) = \sigma$ and $\sigma' \models [ \;]$. By the definition of $\mathsf{unify}$, we know $\sigma = \sigma_\mathbb{E}$. So we must show $\exists \sigma''.\forall \alpha. \sigma' (\alpha) = (\sigma_\mathbb{E} \circ \sigma'')(\alpha)$.
Let  $\sigma'$ be the witness for $\sigma''$ in $\exists \sigma''.\forall \alpha. \sigma' (\alpha) = (\sigma_\mathbb{E} \circ \sigma'')(\alpha)$. Choose an arbitrary $\alpha$. Then we must show $\sigma' (\alpha) = (\sigma_\mathbb{E} \circ \sigma')(\alpha)$. But by Theorem \ref{thm:composition_apply}, we have $(\sigma_\mathbb{E} \circ \sigma')(\alpha) = \sigma' (\sigma_\mathbb{E}(\alpha))$. So we must show 
$\sigma' (\sigma_\mathbb{E}(\alpha)) = \sigma' (\alpha)$. But that follows since $\sigma_\mathbb{E}(\alpha) = \alpha$.
\item Case $C \neq [\;]$. We consider the various cases based on the constraint at the head of the constraint list:
\begin{enumerate}
\item Case $(\alpha \cequal \alpha)::C'$. Apply the induction hypothesis and then this case is trivial.
\item Case $(\alpha \cequal \beta)::C'$ and $\alpha \neq \beta$. Reasoning is similar to case 3 below.  
\item Case $(\alpha \cequal \tau_1 \rightarrow \tau_2)::C'$ and $\alpha \notin \mathsf{FTV}(\tau_1 \rightarrow \tau_2)$.
 We know 
$\mathsf{unify}( \{ \alpha \mapsto \tau_1 \rightarrow \tau_2 \}(C')) = \sigma_1$ and
 the induction hypothesis is \\
$\forall \sigma. \; \forall \sigma'.\; (\mathsf{unify} (\{\alpha \mapsto \tau_1 \rightarrow \tau_2\} (C')) = \sigma\; \wedge \;\sigma' \models (\{\alpha \mapsto \tau_1 \rightarrow \tau_2\} (C')))$\\
$ \mbox{\hspace{1cm}}\Rightarrow \exists \sigma''. \;\forall \alpha'.\; \sigma' (\alpha') = (\sigma \circ \sigma'')(\alpha')$. \\
We  must show \\
$\forall \sigma_p.\; \forall \sigma_2.\; \sigma_p = (\{\alpha \mapsto \tau_1 \rightarrow \tau_2\} \circ \sigma_1) \wedge \sigma_2 \models ((\alpha \cequal \tau_1 \rightarrow \tau_2)::C')$\\
$\mbox{\hspace{1cm}}  \Rightarrow \exists \sigma_3. \;\forall \alpha''.\; \sigma_2(\alpha'')= 
(\sigma_p \circ \sigma_3)(\alpha'')$.\\
Pick an arbitrary $\sigma_p$ and $\sigma_2$. \\Assume $\sigma_p = \{\alpha \mapsto \tau_1 \rightarrow \tau_2\} \circ \sigma_1$ 
and $\sigma_2 \models ((\alpha \cequal \tau_1 \rightarrow \tau_2)::C')$. We must show \\
$\exists \sigma_3. \;\forall \alpha''.\; \sigma_2(\alpha'')= 
((\{\alpha \mapsto \tau_1 \rightarrow \tau_2\} \circ \sigma_1) \circ \sigma_3)(\alpha'')$. Since $\sigma_2 \models ((\alpha \cequal \tau_1 \rightarrow \tau_2)::C')$ so, by the definition of constraint satisfiability, we know $\sigma_2 (\alpha) = \sigma_2(\tau_1 \rightarrow \tau_2)$ and 
$\sigma_2 \models C'$. Then, by Lemma \ref{lemma:constraint_satisfaction_and_substitution_instance} and by our assumptions, we know 
$\sigma_2 \models (\{\alpha \mapsto \tau_1 \rightarrow \tau_2\} (C'))$.
 Since we also know
$\mathsf{unify} (\{\alpha \mapsto \tau_1 \rightarrow \tau_2\} (C')) = \sigma_1$, so, by the induction hypothesis, we know
 $\exists \sigma''. \;\forall \alpha'.\; \sigma_2 (\alpha') = (\sigma_1 \circ \sigma'')(\alpha')$.
We assume  $\forall \alpha'.\; \sigma_2 (\alpha') = (\sigma_1 \circ \sigma_4)(\alpha')$, where $\sigma_4$ is fresh.
Then, to show $\exists \sigma_3. \forall \alpha''. \sigma_2(\alpha'')= 
((\{\alpha \mapsto \tau_1 \rightarrow \tau_2\}\circ \sigma_1)\circ \sigma_3)(\alpha'')$, we choose the witness $\sigma_4$ and show $\forall \alpha''. \sigma_2(\alpha'')= 
((\{\alpha \mapsto \tau_1 \rightarrow \tau_2\}\circ \sigma_1)\circ \sigma_4)(\alpha'')$. Pick an arbitrary $\alpha''$ and show 
$\sigma_2(\alpha'')= 
((\{\alpha \mapsto \tau_1 \rightarrow \tau_2\}\circ \sigma_1)\circ \sigma_4)(\alpha'')$. By Theorem \ref{thm:composition_apply}, we must show \\
$\sigma_2(\alpha'')= \sigma_4(\sigma_1(\{\alpha \mapsto \tau_1 \rightarrow \tau_2\}(\alpha''))$.
There are two cases to consider:
\begin{enumerate}
\item Case $\alpha \neq \alpha''$. Then we must show $\sigma_2(\alpha'')= 
\sigma_4(\sigma_1(\alpha''))$. But that follows our assumptions and Theorem \ref{thm:composition_apply}.
\item Case $\alpha = \alpha''$. Then we must show $\sigma_2(\alpha)= 
\sigma_4(\sigma_1(\tau_1 \rightarrow \tau_2))$. Since we know \\ $\sigma_2(\alpha) = \sigma_2(\tau_1 \rightarrow \tau_2)$, so we must show $\sigma_2(\tau_1 \rightarrow \tau_2)= 
\sigma_4(\sigma_1(\tau_1 \rightarrow \tau_2))$. But that follows from our assumptions and Lemma \ref{lemma:squiggle_ext_lift} and Theorem \ref{thm:composition_apply}.
\end{enumerate}
\item Case $(\tau_1 \rightarrow \tau_2 \cequal \alpha)::C$ and $\alpha \notin \mathsf{FTV}(\tau_1 \rightarrow \tau_2)$. Same as case 3 above.
\item Case $(\tau_1 \rightarrow \tau_2 \cequal \tau_3 \rightarrow \tau_4)::C$. Apply the induction hypothesis and then this case is trivial.
\end{enumerate}
\end{description}
\end{proof}

\subsection{Axiom iii}
\begin{lemma}\label{lemma:compose_and_domain_membership}
$\forall{}\alpha, \alpha'. \:\forall{}\tau. \forall{} \sigma.\: \alpha' \in \mathsf{dom\_subst}(\{\alpha \mapsto \tau\} \circ \sigma)
 \Rightarrow $\\
 $\mbox{\hspace{2.5cm}} \alpha' \in \mathsf{dom\_subst}(\{\alpha \mapsto \tau\})\; \vee \; \alpha' \in \mathsf{dom\_subst}(\sigma)$
\end{lemma}

\begin{lemma}\label{lemma:compose_and_range_membership}
$\forall{}\alpha, \alpha'. \:\forall{}\tau. \:\forall{} \sigma. \: (\alpha \notin \mathsf{FTV}(\tau)\; \wedge\;   \alpha' \in \mathsf{range\_subst}( \{\alpha \mapsto \tau\} \circ \sigma))  \Rightarrow$\\
$\mbox{\hspace{2.5cm}} \alpha' \in \mathsf{range\_subst}(\{\alpha \mapsto \tau\})\; \vee\; \alpha' \in \mathsf{range\_subst}(\sigma)$
\end{lemma}

\noindent
Without going into the details, the following lemma helps us in proving Lemma \ref{lemma:compose_and_range_membership}. Note that the definition of $\circ$ contains references to higher order functions $\mathsf{M.map2}$ and this lemma helps in not having to reason about 
$\mathsf{M.map2}$ function but instead we use Theorem \ref{thm:composition_apply} to reason about substitution composition.
\begin{lemma}\label{lemma:alt_range_def}
$\forall{}\alpha .\: \forall{} \sigma.\; \alpha \in \mathsf{range\_subst}(\sigma) \Leftrightarrow \exists \alpha'. \alpha' \in \mathsf{dom\_subst} (\sigma)\; \wedge\; \alpha \in \mathsf{FTV}(\sigma(\alpha'))  $
\end{lemma}

\begin{lemma}\label{lemma:termination_helper12}
$\forall \alpha, \alpha'. \; \forall \tau. \; \forall C.\; (\alpha' \notin \mathsf{FTV}(\tau) 
\wedge \alpha' \in \mathsf{FTV}(\{\alpha \mapsto \tau\}(C))) \Rightarrow \alpha' \in \mathsf{FTV}(C)$.
\end{lemma}

\begin{lemma}\label{lemma:mgu_axiom3_helper1}
$\forall{} C.\;\forall{}\sigma. \;\mathsf{unify}(C) = \sigma \:  \Rightarrow \mathsf{dom\_subst}(\sigma) \subseteq \mathsf{FTV}(C)$
\end{lemma}
\begin{proof}
By functional induction on $\mathsf{unify}(C)$ and Lemma \ref{lemma:compose_and_domain_membership}.
\end{proof}

\noindent
We focus on the proof of the most involved lemma.
\begin{lemma}\label{lemma:mgu_axiom3_helper2}
$\forall{} C.\;\forall{}\sigma. \;\mathsf{unify}(C) = \sigma \:  \Rightarrow \mathsf{range\_subst}(\sigma) \subseteq \mathsf{FTV}(C)$
\end{lemma}
\begin{proof}
Choose an arbitrary $C$. Unfolding the definition of $\subseteq$, we must
show\\ $\forall{}\sigma. \;\mathsf{unify}(C) = \sigma \: \Rightarrow \forall
\alpha'. \;\alpha' \in \mathsf{range\_subst}(\sigma) \Rightarrow \alpha' \in
\mathsf{FTV}(C)$. By functional induction on $\mathsf{unify}(C)$, there are two
main cases:
\begin{description}
\item Case $C = [\;]$. Then, by the definition of $\mathsf{unify}$, we know $\sigma = \sigma_\mathbb{E}$. So we must show \\
$\mathsf{range\_subst}(\sigma_\mathbb{E}) \subseteq \mathsf{FTV}(\emptylist)$. The proof follows from  the definition of $\mathsf{range\_subst}$ and the definition of $\mathsf{FTV}$.
\item Case $C \neq [\;]$. We consider the various cases based on the constraint at the head of the constraint list:
\begin{enumerate}
\item Case $(\alpha \cequal \alpha)::C'$.  The induction hypothesis is:\\
$ \forall{}\sigma. \;\mathsf{unify}(C') = \sigma \:  \Rightarrow \forall \alpha''. \alpha'' \in \mathsf{range\_subst}(\sigma) \Rightarrow \alpha'' \in \mathsf{FTV}(C')$\\
and we must show \\
$\forall{}\sigma. \;\mathsf{unify}(C') = \sigma \:  \Rightarrow \forall \alpha'. \alpha' \in \mathsf{range\_subst}(\sigma) \Rightarrow \alpha' \in \mathsf{FTV}((\alpha \cequal \alpha)::C')$. \\
Pick an arbitrary $\sigma$ and assume $\mathsf{unify}(C') = \sigma$. Pick an arbitrary $\alpha'$. \\
Assume $\alpha' \in \mathsf{range\_subst}(\sigma)$ and show $\alpha' \in \mathsf{FTV}((\alpha \cequal \alpha)::C')$.\\
Since we know $\mathsf{unify}(C') = \sigma$, so, by the induction hypothesis, we know \\
$\forall \alpha''. \; \alpha'' \in \mathsf{range\_subst}(\sigma) \Rightarrow \alpha'' \in \mathsf{FTV}(C')$. Since we also know $\alpha' \in \mathsf{range\_subst}(\sigma)$, so we know $\alpha' \in \mathsf{FTV}(C')$. That also means $\alpha' \in \mathsf{FTV}((\alpha \cequal \alpha)::C')$ as was to be shown.

\item Case $(\alpha \cequal \beta)::C'$ and $\alpha \neq \beta$. Reasoning is similar to case 3 below.  
\item Case $(\alpha \cequal \tau_1 \rightarrow \tau_2)::C'$ and $\alpha \notin \mathsf{FTV}(\tau_1 \rightarrow \tau_2)$.
 We know 
$\mathsf{unify} (\{\alpha \mapsto \tau_1 \rightarrow \tau_2\} (C')) = \sigma_1$, and
 the induction hypothesis is \\
$\; \forall \sigma'.\; \mathsf{unify}(\{\alpha \mapsto \tau_1 \rightarrow \tau_2\}(C')) = \sigma' \Rightarrow$\\  
$\mbox{\hspace{1cm}}\forall \alpha'. \alpha' \in \mathsf{range\_subst}(\sigma') \Rightarrow \alpha' \in \mathsf{FTV}(\{\alpha \mapsto \tau_1 \rightarrow \tau_2\}(C')) $.\\ 
We  must show \\
$ \forall \alpha''. \alpha'' \in \mathsf{range\_subst}(\{\alpha \mapsto \tau_1 \rightarrow \tau_2\} \circ \sigma_1) \Rightarrow \alpha'' \in \mathsf{FTV}((\alpha \cequal \tau_1 \rightarrow \tau_2)::C')$.\\
Pick an arbitrary $\alpha''$ and assume  $\alpha'' \in \mathsf{range\_subst}(\{\alpha \mapsto \tau_1 \rightarrow \tau_2\} \circ \sigma_1)$.   We must show $\alpha'' \in \mathsf{FTV}(\{\alpha \cequal \tau_1 \rightarrow \tau_2\}::C')$. There are two cases:
\begin{enumerate}
\item Case $\alpha'' = \alpha$. Then clearly $\alpha'' \in \mathsf{FTV}(\{\alpha \cequal \tau_1 \rightarrow \tau_2\}(C'))$ as was to be shown.
\item Case $\alpha'' \neq \alpha$.  Then we have two cases:
    \begin{enumerate}
      \item $\alpha'' \in \mathsf{FTV}(\tau_1 \rightarrow \tau_2)$. Then clearly $\alpha'' \in \mathsf{FTV}(\{\alpha \cequal \tau_1 \rightarrow \tau_2\}::C')$ as was to be shown.
      \item $\alpha'' \notin \mathsf{FTV}(\tau_1 \rightarrow \tau_2)$. Then we must show $\alpha'' \in \mathsf{FTV}(C')$. Since we know\\ $\mathsf{unify} (\{\alpha \mapsto \tau_1 \rightarrow \tau_2\} (C')) = \sigma_1$, so by the induction hypothesis we know \\
$\forall \alpha'. \alpha' \in \mathsf{range\_subst}(\sigma_1) \Rightarrow \alpha' \in \mathsf{FTV}(\{\alpha \mapsto \tau_1 \rightarrow \tau_2\}(C')) $.\\ Since $\alpha'' \in \mathsf{range\_subst}(\{\alpha \mapsto \tau_1 \rightarrow \tau_2\} \circ \sigma_1)$ so, by Lemma \ref{lemma:compose_and_range_membership}, we know either $\alpha'' \in \mathsf{range\_subst}(\{\alpha \mapsto \tau_1 \rightarrow \tau_2\})$ or $\alpha'' \in \mathsf{range\_subst}(\sigma_1)$. Again, there are two cases:
       \begin{enumerate}
       \item Case $\alpha'' \in \mathsf{range\_subst}(\{\alpha \mapsto \tau_1 \rightarrow \tau_2\})$. Then $\alpha'' \in \mathsf{FTV}(\tau_1 \rightarrow \tau_2)$ - a contradiction. 
       \item Case $\alpha'' \in \mathsf{range\_subst}(\sigma_1)$. Then from the induction hypothesis we know\\ $\alpha'' \in \mathsf{FTV}(\{\alpha \mapsto \tau_1 \rightarrow \tau_2\}(C')) $. Then by Lemma \ref{lemma:termination_helper12}, $\alpha'' \in \mathsf{FTV}(C')$ as was to be shown.
       \end{enumerate}
    \end{enumerate}
\end{enumerate}

\item Case $(\tau_1 \rightarrow \tau_2 \cequal \alpha)::C$ and $\alpha \notin \mathsf{FTV}(\tau_1 \rightarrow \tau_2)$.
Same as case 3 above.
\item Case $(\tau_1 \rightarrow \tau_2 \cequal \tau_3 \rightarrow \tau_4)::C$. Apply the induction hypothesis and then this case is trivial.
\end{enumerate}
\end{description}
\end{proof}


\begin{theorem}\label{thm:mgu_axiom3}
$\forall{} C.\;\forall{}\sigma. \;\mathsf{unify}\: C = \sigma \:  \Rightarrow \mathsf{FTV}(\sigma) \subseteq \mathsf{FTV}(C)$
\end{theorem}
\begin{proof}
By the definition of $\mathsf{FTV}$  and  by Lemma \ref{lemma:mgu_axiom3_helper1} and Lemma \ref{lemma:mgu_axiom3_helper2}.
\end{proof}

\subsection{Axiom iv}
This axiom requires the notion of subterms, which we define below:\\

 \begin{tabular}{lcl}
     $\mbox{\hspace{1cm}}\mathsf{subterms}(\alpha)$ & $\definedAs$ & [\;]\\
     $\mbox{\hspace{1cm}}\mathsf{subterms}(\tau_1 \rightarrow \tau_2)$ &$\definedAs$ &$\tau_1::\tau_2::(\mathsf{subterms} \; \tau_1)\,\, \mathrm{++}\,\, (\mathsf{subterms}  \; \tau_2)$ \\\\
\end{tabular}

\noindent
Then we can define what it means to for a term to be contained in another term.
\begin{lemma}\label{lemma:containment}
$\forall{}\tau,\tau'.\:\tau \in \mathsf{subterms}(\tau') \:\Rightarrow \: \forall{}\tau''.\:  \tau'' \in \mathsf{subterms}(\tau) \Rightarrow  \tau'' \in \mathsf{subterms}(\tau')  $
\end{lemma}

\noindent
A somewhat related lemma is used to show well foundedness of types.
\begin{lemma}\label{lemma:well_foundedness_for_types}
$\forall{}\tau.\:\neg\: \tau \in \mathsf{subterms}(\tau) $
\end{lemma}
\begin{proof}
By induction on the structure of the type $\tau$ and by Lemma \ref{lemma:containment}.
\end{proof}

\noindent
The following obvious but powerful lemma helps in proving the axiom:
\begin{lemma}\label{lemma:member_subterms_and_apply_subst}
$\forall{} \sigma.\: \forall{} \alpha.\: \forall{}\tau.\: \alpha \in \mathsf{subterms}(\tau) \Rightarrow \sigma(\alpha) \neq \sigma(\tau) $
\end{lemma}
\begin{proof}
By induction on the structure of the type $\tau$ and by Lemma \ref{lemma:well_foundedness_for_types}.
\end{proof}

\begin{lemma}\label{lemma:member_arrow_and_subterms}
$\forall{} \sigma.\: \forall{}\alpha.\: \forall{}\tau_1, \tau_2.\: \alpha \in  \mathsf{FTV}(\tau_1) \vee  \alpha \in \mathsf{FTV}(\tau_2) \Rightarrow \alpha \in \mathsf{subterms}(\tau_1 \rightarrow \tau_2) $
\end{lemma}
\begin{proof}
By induction on $\tau_1$, followed by induction on $\tau_2$.
\end{proof}

\noindent
A corollary from the above two gives us the required lemma.
\begin{corollary}\label{lemma:member_apply_subst_unequal}
$\forall{} \sigma.\: \forall{} \alpha.\: \forall{}\tau_1, \tau_2.\:\alpha  \in \mathsf{FTV}(\tau_1) \vee  \alpha  \in \mathsf{FTV}(\tau_2) \Rightarrow \sigma (\alpha) \neq \sigma(\tau_1 \rightarrow \tau_2) $
\end{corollary}
\begin{proof}
By Lemma \ref{lemma:member_subterms_and_apply_subst} and \ref{lemma:member_arrow_and_subterms}.
\end{proof}

This is the only theorem where the failure cases are interesting. So in the following theorem we carry along the constructor that shows success or failure of $\mathsf{unify}$ function call.
\begin{theorem}
$ \forall{} C. \; \forall{}\sigma.\; \sigma \models  C  \:  \Rightarrow \exists \sigma'.\; \mathsf{unify}(C) = \mathsf{Some} \;\sigma'$
\end{theorem}
\begin{proof}
Choose an arbitrary $C$ and $\sigma$.  By functional induction on $\mathsf{unify}(C)$, there are two main cases: 
\begin{description}
\item Case $C = [\;]$. Assume $\sigma \models \emptylist$. Then we must show $\exists \sigma'.\; \mathsf{unify}(\emptylist) = \mathsf{Some} \;\sigma'$. Let $\sigma_\mathbb{E}$  be the witness for $\sigma'$ in $\exists \sigma'.\; \mathsf{unify}(\emptylist) = \mathsf{Some} \;\sigma'$. So we must show $\mathsf{unify}\; \emptylist = \mathsf{Some}\; \sigma_\mathbb{E}$ but that follows from the definition of $\mathsf{unify}$.
\item Case $C \neq [\;]$. We consider the various cases based on the constraint at the head of the constraint list:
\begin{enumerate}
\item Case $(\alpha \cequal \alpha)::C'$. Apply the induction hypothesis and then this case is trivial.
\item Case $(\alpha \cequal \beta)::C'$ and $\alpha \neq \beta$. Reasoning is similar to case 3 below.  
\item Case $(\alpha \cequal \tau_1 \rightarrow \tau_2)::C'$ and $\alpha \notin \mathsf{FTV}(\tau_1 \rightarrow \tau_2)$.
 We know \\
 $\mathsf{unify} (\{\alpha \mapsto \tau_1 \rightarrow \tau_2\} (C')) = \mathsf{None}$ and
 the induction hypothesis is: \\
$\sigma' \models (\{\alpha \mapsto \tau_1 \rightarrow \tau_2\} (C')) \Rightarrow \exists \sigma''.\; \mathsf{unify} (\{\alpha \mapsto \tau_1 \rightarrow \tau_2\} (C')) = \mathsf{Some}\;\sigma''$.\\
We  must show \\
$\sigma' \models ((\alpha \cequal \tau_1 \rightarrow \tau_2)::C')  \Rightarrow 
\exists \sigma_3.\; \mathsf{None} = \mathsf{Some}\; \sigma_3$.\\
Assume $\sigma' \models ((\alpha \cequal \tau_1 \rightarrow \tau_2)::C')$, {\em i.e.}, $\sigma'(\alpha) = \sigma'(\tau_1 \rightarrow \tau_2)$ and 
$\sigma' \models C'$. \\
We  must show $\exists \sigma_3.\; \mathsf{None} = \mathsf{Some}\; \sigma_3$.
By Lemma \ref{lemma:constraint_satisfaction_and_substitution_instance} and by our assumptions, we know\\ 
$\sigma' \models (\{\alpha \mapsto \tau_1 \rightarrow \tau_2\} (C'))$. So, by the induction hypothesis, we know $\exists \sigma''. \; \mathsf{None} = \mathsf{Some}\;\sigma''$. Since  we know $\exists \sigma''. \; \mathsf{None} = \mathsf{Some}\;\sigma''$, so assume 
$\mathsf{None} = \mathsf{Some}\;\sigma'''$, where $\sigma'''$ is fresh, but that is a contradiction and so this case holds.
\item Case $(\alpha \cequal \tau_1 \rightarrow \tau_2)::C'$ and $\alpha \in \mathsf{FTV}(\tau_1 \rightarrow \tau_2)$.\\
Then, we must show $\sigma' \models ((\alpha \cequal \tau_1 \rightarrow \tau_2)::C')  \Rightarrow 
\exists \sigma_3. \mathsf{None} = \mathsf{Some}\; \sigma_3$.\\
 Assume $\sigma' \models ((\alpha \cequal \tau_1 \rightarrow \tau_2)::C')$, {\em i.e.}, $\sigma' (\alpha) = \sigma'(\tau_1 \rightarrow \tau_2)$ and $\sigma' \models C'$. Since we know 
$\alpha \in \mathsf{FTV}(\tau_1 \rightarrow \tau_2)$, {\em i.e.},  either $\alpha \in \mathsf{FTV}(\tau_1)$ or  $\alpha \in \mathsf{FTV}(\tau_2)$,
 so by Corollary \ref{lemma:member_apply_subst_unequal}\\ $\sigma' (\alpha) \neq \sigma'(\tau_1 \rightarrow \tau_2)$, which is a contradiction. Thus the proof follows trivially.
\item Case $(\tau_1 \rightarrow \tau_2 \cequal \alpha)::C$ and $\alpha \notin \mathsf{FTV}(\tau_1 \rightarrow \tau_2)$.
Similar to case 3.
\item Case $(\tau_1 \rightarrow \tau_2 \cequal \alpha)::C$ and $\alpha \in \mathsf{FTV}(\tau_1 \rightarrow \tau_2)$.
Similar to case 4.
\item Case $(\tau_1 \rightarrow \tau_2 \cequal \tau_3 \rightarrow \tau_4)::C$. Apply the induction hypothesis and then this case is trivial.
\end{enumerate}
\end{description}
\end{proof}



\subsection{Axiom v}
The following lemmas are needed for the main proof, the first two follow by induction on the structure of the type $\tau$ and the third by induction on $C$.
\begin{lemma}\label{idempotent_helper_helper1_gen}
$\forall \sigma.\; \forall \alpha. \;\forall \tau.\; \alpha \notin \mathsf{FTV}(\tau)  \wedge \alpha \notin \mathsf{FTV}(\sigma) \Rightarrow \alpha \notin \mathsf{FTV}\;(\sigma (\tau))$
\end{lemma}

\begin{lemma}\label{fresh_type_and_member_converse}
$\forall \alpha.\; \forall \tau, \tau'. \;\alpha \notin \mathsf{FTV}(\tau)  \Rightarrow \{\alpha \mapsto \tau'\} (\tau)= \tau$
\end{lemma}

\begin{lemma}\label{lemma:apply_subst_constr_general}
$\forall \alpha.\; \forall{}\tau.\; \forall{} C. \;\alpha \notin \mathsf{FTV}(\tau) \Rightarrow \alpha \notin \mathsf{FTV} (\{\alpha \mapsto \tau \}(C))$
\end{lemma}

\noindent
The theorem we must prove is:
\begin{theorem}
$\forall{} C.\; \forall{}\sigma.\;  \mathsf{unify}(C)= \sigma \:  \Rightarrow  (\sigma \circ \sigma) \approx \sigma$.
\end{theorem}
\begin{proof}
Pick an arbitrary $C$. Unfolding the definition of $\approx$, and by Theorem \ref{thm:composition_apply}, we must show:\\
$\forall{}\sigma.\; \mathsf{unify}(C) = \sigma \Rightarrow \forall \alpha.\; \sigma (\sigma (\alpha))= \sigma (\alpha)$.\\
 By functional induction on $\mathsf{unify}\; C$, there are two main cases: 
\begin{description}
\item Case $C = [\;]$. 
 This case follows since $\forall \alpha. \;\sigma_\mathbb{E} (\alpha) = \alpha$.
\item Case $C \neq [\;]$. We consider the various cases based on the constraint at the head of the constraint list:
\begin{enumerate}
\item Case $(\alpha \cequal \alpha)::C'$. Apply the induction hypothesis and then this case is trivial.
\item Case $(\alpha \cequal \beta)::C'$. Reasoning is similar to case 3 below.
\item Case $(\alpha \cequal \tau_1 \rightarrow \tau_2)::C'$ and $\alpha \notin \mathsf{FTV}(\tau_1 \rightarrow \tau_2)$. We know $\mathsf{unify}\; (\{\alpha \mapsto \tau_1 \rightarrow \tau_2\}(C'))= \sigma$ and the induction hypothesis is:\\
$\forall \sigma'. \mathsf{unify}\; \{\alpha \mapsto \tau_1 \rightarrow \tau_2\}(C')=  \sigma' \Rightarrow \forall \alpha'. \sigma' (\alpha') = \sigma'(\sigma' (\alpha'))$\\
And we must show:\\
$ \sigma (\{\alpha \mapsto \tau_1 \rightarrow \tau_2 \}(\alpha'')) = (\sigma(\{\alpha \mapsto \tau_1 \rightarrow \tau_2 \}(\sigma(\{\alpha \mapsto \tau_1 \rightarrow \tau_2 \}(\alpha'')))))$.\\
There are two cases:
\begin{enumerate}
\item Case $\alpha  = \alpha''$. Then we must show $\sigma(\tau_1 \rightarrow \tau_2) = \sigma(\{\alpha \mapsto \tau_1 \rightarrow \tau_2 \}(\sigma(\tau_1 \rightarrow \tau_2)))$. From Lemma \ref{lemma:apply_subst_constr_general} and Theorem \ref{thm:mgu_axiom3}, we know that $\alpha \notin \mathsf{FTV}(\sigma)$.
Since $\alpha \notin \mathsf{FTV}(\tau_1 \rightarrow \tau_2)$ and $\alpha \notin \mathsf{FTV}(\sigma)$, so by Lemma \ref{idempotent_helper_helper1_gen}, 
$\alpha \notin \mathsf{FTV}(\sigma(\tau_1 \rightarrow \tau_2))$. By Lemma \ref{fresh_type_and_member_converse} (choosing $\tau'$ to be $\tau_1 \rightarrow \tau_2$), we get $\sigma(\tau_1 \rightarrow \tau_2) =  \{\alpha \mapsto \tau_1 \rightarrow \tau_2\}(\sigma (\tau_1 \rightarrow \tau_2))$. So now we must show 
$\sigma (\tau_1 \rightarrow \tau_2) =  \sigma(\sigma  (\tau_1 \rightarrow \tau_2))$. Then, by Lemma \ref{lemma:squiggle_ext_lift}, we must show 
$\forall \beta. \;\sigma (\beta) =  \sigma(\sigma  (\beta))$ . Choose an arbitrary $\beta$ and show $\sigma (\beta) =  \sigma(\sigma  (\beta))$, but that follows from the induction hypothesis (by choosing $\sigma'$ to be $\sigma$ and $\alpha'$ to be $\beta$) and our assumptions.
\item Case $\alpha \neq \alpha''$. Then we must show $\sigma(\alpha'') = \sigma(\{\alpha \mapsto \tau_1 \rightarrow \tau_2 \}(\sigma(\alpha'')))$. From Lemma \ref{lemma:apply_subst_constr_general} and Theorem \ref{thm:mgu_axiom3}, we know that $\alpha \notin \mathsf{FTV}(\sigma)$.
Since $\alpha \notin \mathsf{FTV}(\alpha'')$ and $\alpha \notin \mathsf{FTV}(\sigma)$, so by Lemma \ref{idempotent_helper_helper1_gen}, 
$\alpha \notin \mathsf{FTV}(\sigma(\alpha''))$. By Lemma \ref{fresh_type_and_member_converse} and using $\tau'$ to be $\tau_1 \rightarrow \tau_2$ we get $\sigma(\alpha'') =  (\{\alpha \mapsto \tau_1 \rightarrow \tau_2\}(\sigma (\alpha'')))$. So now we must show 
$\sigma (\alpha'') =  \sigma(\sigma  (\alpha''))$, but that follows from the induction hypothesis (by choosing $\sigma'$ to be $\sigma$ and $\alpha'$ to be $\alpha''$) and our assumptions. 

 \end{enumerate}
\item Case $(\tau_1 \rightarrow \tau_2 \cequal \alpha)::C'$ and $\alpha \notin \mathsf{FTV}(\tau_1 \rightarrow \tau_2)$. Same as Case 3.
\item Case $(\tau_1 \rightarrow \tau_2 \cequal \tau_3 \rightarrow \tau_4)::C'$. Apply the induction hypothesis and then this case is trivial.
\end{enumerate}
\end{description}
\end{proof}


\subsection{Axiom vi}
The theorem we must prove is:
\begin{theorem}\label{thm:empty_map_axiom}
$\forall \sigma. \mathsf{unify}\;[\;] = \sigma \Rightarrow \;\sigma =  \sigma_\mathbb{E} $
\end{theorem}  
\begin{proof}
Choose an arbitrary $\sigma$. Assume  $\mathsf{unify}\;[\;] = \sigma$. Unfold the definition of $\mathsf{unify}$.  Then we know $\sigma = \sigma_\mathbb{E}$ as was to be shown.
\end{proof}

\subsection{Axiom vii}
The main proof requires a lemma, which we mention next.
\begin{lemma}\label{lemma:unify_and_append_helper2}
$\forall C, C'. \;\forall \alpha.\; \forall \tau.\; \{\alpha \mapsto \tau\}(C)\; {++}\; \{\alpha \mapsto \tau\}(\mathbb{C'}) = \{\alpha \mapsto \tau\} (C \;{++}\; C')$ 
\end{lemma}

\noindent 
The theorem we must prove is:
\begin{theorem}
$\forall C, C_2. \;\forall \sigma', \sigma'', \sigma'''. \;(\mathsf{unify}(C) = \sigma'\: \wedge \: \mathsf{unify}(\sigma'(C_2)) = \sigma''\; \wedge\; \mathsf{unify}(C\; {++}\; C_2) = \sigma''')$\\
 \mbox{\hspace{2cm}}$\Rightarrow  \;\sigma''' \approx (\sigma' \circ \sigma'')$
\end{theorem}
\begin{proof}
Pick an arbitrary $C$. By Theorem \ref{thm:composition_apply} and unfolding the definition of $\approx$, we must show:\\
$\forall C_2. \;\forall \sigma', \sigma'', \sigma'''. \;(\mathsf{unify}(C) = \sigma'\: \wedge \: \mathsf{unify}(\sigma'(C_2)) = \sigma'' \: \wedge \: \mathsf{unify}(C\: \mathit{++}\: C_2) =  \sigma''')$\\
 $\mbox{\hspace{3cm}}  \Rightarrow \forall \alpha'. \sigma'''(\alpha') = \sigma'' (\sigma' (\alpha'))$.\\
By functional induction on $\mathsf{unify}(C)$, there are two main cases:
\begin{description}
\item Case $C = [\;]$. Follows from Theorem \ref{thm:empty_map_axiom} and the assumptions.
\item Case $C \neq [\;]$. 
Consider the various cases based on the constraint at the head of the constraint list.
\begin{enumerate}
\item Case $(\alpha \cequal \alpha)::C'$. This case follows from the induction hypothesis and the definition of append.
\item Case $(\alpha \cequal \beta)::C'$ and $ \alpha \neq \beta$. Similar to case 3 below. 
\item Case $(\alpha \cequal \tau_1 \rightarrow \tau_2)::C'$ and $\alpha \notin \mathsf{FTV}(\tau_1 \rightarrow \tau_2)$.
We know $\mathsf{unify}(\{\alpha \mapsto \tau_1 \rightarrow \tau_2\}(C')) = \sigma$.
The induction hypothesis is:\\
$\forall C_1.\; \forall \sigma_1, \sigma_2, \sigma_3.$\\
$\mbox{\hspace{1cm}} \;(\mathsf{unify} (\{\alpha \mapsto \tau_1 \rightarrow \tau_2\}(C')) = \sigma_1 \; \wedge$\\
$\mbox{\hspace{1.25cm}}  \mathsf{unify}(\sigma_1(C_1)) = \sigma_2 \; \wedge $\\
$\mbox{\hspace{1.25cm}}\mathsf{unify} ((\{\alpha \mapsto \tau_1 \rightarrow \tau_2\}(C')) {++} C_1) = \sigma_3)$\\
  $\mbox{\hspace{2cm}}\Rightarrow \forall \alpha''.\; \sigma_3 (\alpha'') = \sigma_2 (\sigma_1 (\alpha''))$.\\
We must show:\\
$\forall C_2.\; \forall \sigma', \sigma'', \sigma'''.$\\
 $\mbox{\hspace{1cm}}(\{\alpha \mapsto \tau_1 \rightarrow \tau_2\}\circ \sigma = \sigma' \wedge$\\
$\mbox{\hspace{1.25cm}}\mathsf{unify} (\sigma' ( C_2)) = \sigma'' \wedge$\\
$\mbox{\hspace{1.25cm}}\mathsf{unify} (((\alpha \cequal \tau_1 \rightarrow \tau_2)::C') {++} C_2) = \sigma''')$\\
$\mbox{\hspace{2cm}}\Rightarrow \forall \alpha'. \; \sigma'''(\alpha') = \sigma'' (\sigma' (\alpha'))$.\\
Pick an arbitrary $C_2, \sigma', \sigma''$ and $\sigma'''$.
Assume 
$\{\alpha \mapsto \tau_1 \rightarrow \tau_2\}\circ \sigma = \sigma'$ and \\
$\mathsf{unify}(\sigma'( C_2)) = \sigma''$ and 
$\mathsf{unify}(((\alpha \cequal \tau_1 \rightarrow \tau_2)::C') {++} C_2) = \sigma'''$. 
By the definition of append, the last assumption is $\mathsf{unify} ((\alpha \cequal \tau_1 \rightarrow \tau_2)::(C' {++} C_2)) = \sigma'''$.\\ Unfolding the $\mathsf{unify}$ definition once, we know 
$\mathsf{unify} (\{\alpha \mapsto \tau_1 \rightarrow \tau_2\}(C'\:\: {++}\:\: C_2)) = {\sigma}_T$, where 
 $\sigma''' = \{\alpha \mapsto \tau_1 \rightarrow \tau_2\} \circ \sigma_T$. Also, since $\sigma' = \{\alpha \mapsto \tau_1 \rightarrow \tau_2\}\circ \sigma$, 
  so we know $\mathsf{unify}((\{\alpha \mapsto \tau_1 \rightarrow \tau_2\}\circ \sigma)(C_2)) = \sigma''$.
Since we know $\sigma''' = \{\alpha \mapsto \tau_1 \rightarrow \tau_2\} \circ \sigma_T$, so 
we must show $\forall \alpha'. \;  (\{\alpha \mapsto \tau_1 \rightarrow \tau_2\} \circ \sigma_T)(\alpha') = \sigma''((\{\alpha \mapsto \tau_1 \rightarrow \tau_2\}\circ \sigma)(\alpha'))$.
Pick an arbitrary $\alpha'$. By Theorem \ref{thm:composition_apply}, we must show \\
  $\sigma_T(\{\alpha \mapsto \tau_1 \rightarrow \tau_2\} (\alpha')) = \sigma'' (\sigma (\{\alpha \mapsto \tau_1 \rightarrow \tau_2\}(\alpha')))$.
There are two cases:
\begin{enumerate}
\item Case $\alpha = \alpha'$.  Then we must show $\sigma_T(\tau_1 \rightarrow \tau_2) = \sigma''(\sigma(\tau_1 \rightarrow \tau_2))$.  But by Lemma \ref{lemma:squiggle_ext_lift}, we must show
 $\forall \alpha'''.\;\sigma_T(\alpha''') = \sigma''(\sigma(\alpha'''))$. Pick an arbitrary $\alpha'''$ and so we must show 
 $\sigma_T(\alpha''') = \sigma''(\sigma(\alpha'''))$.
  But that follows from the induction hypothesis (by choosing $C_1$ to be $\{\alpha \mapsto \tau_1 \rightarrow \tau_2\}(C_2)$, $\sigma_1$ to be $\sigma$, $\sigma_2$ to be $\sigma''$ and $\sigma_3$ to be $\sigma_T$) and the definition of substitution composition
 and Lemma \ref{lemma:unify_and_append_helper2} and the assumptions.

\item Case $\alpha \neq \alpha' $.
Then we must show $\sigma_T(\alpha') = \sigma''(\sigma(\alpha'))$.  
  But that follows from the induction hypothesis (by choosing $C_1$ to be $\{\alpha \mapsto \tau_1 \rightarrow \tau_2\}(C_2)$, $\sigma_1$ to be $\sigma$, $\sigma_2$ to be $\sigma''$ and $\sigma_3$ to be $\sigma_T$) and the definition of substitution composition
 and Lemma \ref{lemma:unify_and_append_helper2} and the assumptions.

\end{enumerate} 
\item Case $(\tau_1 \rightarrow \tau_2 \cequal \alpha)::C$ and $\alpha \notin \mathsf{FTV}(\tau_1 \rightarrow \tau_2)$.
Same as the above case.
\item Case $(\tau_1 \rightarrow  \tau_2 \cequal \tau_3 \rightarrow \tau_4 )::C$.
Apply the induction hypothesis.
\end{enumerate}
\end{description}
\end{proof}

\section{Related Work and Conclusions}
\label{sec:conclusions}
Unification is
fundamentally used in type inference. There are formalizations of the unification algorithm in a number of different
theorem provers \cite{Blanqui_unification, paper:paulsonLCF, phdthesis:rouyer}.  We comment on the implementation in the CoLoR library \cite{paper:color}. CoLoR
is an extensive and very successful library supporting reasoning about
termination and rewriting.  A Coq implementation of the unification
algorithm was recently released \cite{Blanqui_unification}.  Our implementation
differs from theirs in a number of ways.  Perhaps the most significant
difference is that we represent substitutions as finite maps, whereas in CoLoR the substitutions are represented
 by functions from type variables to a
generalized term structure.  The axioms verified here are not explicitly
verified in CoLoR, however their library could serve as a basis for doing
so.  We believe that the lemmas supporting our verification could be translated
into their more general framework but that the proofs would be significantly
different because we use functional induction which follows the structure of our
algorithm.  The unification algorithm in CoLoR is specified in a significantly different
style (as an iterated step function).  


Though many lemmas were simple, many others required generalization in order
for the proof to go through. Our choice of finite maps library to represent
substitutions helped us significantly. Coq's finite maps library is expressive
enough to specify complicated definitions (substitution composition, range
elements) yet the reasoning with them is simple if we abstract away from the
actual definition and look at the extensional behavior instead.  Since we used
an interface, we could not really argue about the normal substitution equality.
Our specification of unification was in a functional style but the definition
was general recursive. This meant that we had to show the termination using a
well-founded ordering. Once termination was established, the {\tt functional
induction} tactic helped us immensely in reasoning about the first-order
unification algorithm.

The entire formalization (all seven axioms) is
done in Coq 8.1.pl3 version in around 5000 lines of specifications and tactics,
and is available online at \url{http://www.cs.uwyo.edu/~skothari}.

We would like to thank Santiago Zanella (INRIA - Sophia Antipolis) for showing
us how to encode lexicographic ordering for 3-tuples in Coq. We thank 
Frederic Blanqui for answering our queries regarding the new release of CoLoR
library,  Laurent Th\'ery for making his Coq
formulation of Sudoku \cite{paper:sudoku} available on the web, St\'ephane Lescuyer and other Coq-club members 
for answering our queries on the Coq-club mailing list,  and Christian Urban (TU Munich) for discussing at length the MGU axioms used in their verification of Algorithm W \cite{bookchapter:urbannipkow}. Finally, we  want to thank anonymous referees for their detailed comments and 
suggestions (on an earlier draft of this paper), which greatly improved the presentation of this paper.

\bibliography{references}
\bibliographystyle{eptcs}

\end{document}